# Open Heavy Flavor Production in p+p, d+Au and Au+Au Collisions at $\sqrt{s_{NN}} = 200 GeV$

X. R. Wang, for the PHENIX collaboration

*Department of Physics, New Mexico State University, Las Cruces, NM88003*

**Abstract:** PHENIX results for open charm production from semi-lepton decay in p+p, d+Au and Au+Au collisions in a wide rapidity ranges at $\sqrt{s_{NN}} = 200 GeV$ are presented.

## INTRODUCTION

Heavy quarks are a unique tool in High Energy and Nuclear Physics. The charm quark has attracted the attention of both experimentalists and theoreticians since it was discovered in 1974.

The main goal of the RHIC program is the identification and study of the hot high-density matter created in heavy-ion collisions, i.e. the Quark Gluon Plasma (QGP). Heavy quarks, with masses greater than 1 GeV) are widely recognized as the cleanest probes in the laboratory quest for the QGP and their analysis is the vital next step to clinch its discovery. Heavy quarks are not present as constituents of the colliding nuclei, but are formed at the earliest times after the collision. Once formed, heavy quarks live much longer ($\sim 10^{-11}$ sec) than the duration of the QGP ($\sim 10^{-22}$ sec), and travel macroscopic distances (up to a few mm) away from the creation point. This separation of mass scales and time scales allows sampling of the properties of the plasma in a way not possible with light quark.

The PHENIX experiment [1] at the Relativistic Heavy Ion Collider (RHIC) has measured open charm through semi-lepton decay at mid-rapidity $|\eta| < 0.35$ and forward and backward rapidities $1.2 < |\eta| < 2.4$. A study of heavy flavor production in different collision systems in various kinematic regions presents an opportunity to probe cold nuclear medium in p+p, d+Au collisions and hot dense matter effects in Au+Au collisions. Detailed comparison of open charm production in those collisions will help us to answer the fundamental questions: charm production mechanism, cold nuclear effect and lead us to understand the hot density matter created in heavy ion collisions at RHIC.

## EXPERIMENTAL TECHNIQUE

PHENIX Detector was optimized to measure the leptons. A pair of central spectrometers at mid- rapidity to measure electrons, hadrons, and photons, and a pair of muon spectrometers to measure muons[1] in the forward and backward rapidity. How to extract the non-photonic electron and prompt muon from heavy flavor decay form the inclusive single lepton is a very challenge task.

PHENIX has developed robust and accurate methods [2,3,4] of measuring the heavy quark production by disentangling the electron contribution from the semi-leptonic decays of the open charm/bottom particles referred to as "non-photonic" electrons from the electrons created from the conventional sources (Dalitz decays of light mesons, photon-conversions in the detector material - referred to as ``photonic" electrons).

The main background sources of muon from heavy flavor decay (prompt muons) are those from light hadron decay and small fraction of hadron punch through contamination. Since the muon from light hadron decay are collision vertex dependent, it can be subtracted by collision vertex distribution analysis. The data driven method was used to subtract the hadron punch through. By analyzing the hadron stopped at gap 2 and 3, one can estimate the hadron contribution at last gap using the attenuation model [5].

## RESULTS AND SUMMARY

Because of the high charm quark mass, one expect to calculate open charm production rate using perturbative Quantum Chromody-namics (pQCD). Comparing open charm production in p+p collisions with NLO pQCD calculation shows the theory under predicts the lepton production in p-p collisions. See Figure 1 shows $p_T$ spectrum of the single electron (left) and single muon (right) from heavy flavor semi-lepton decay in p+p collisions.

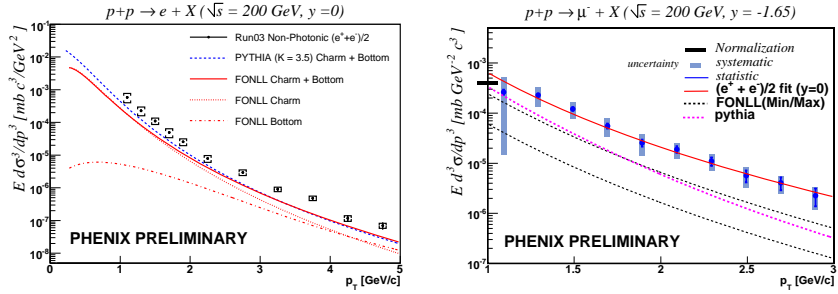

**Figure 1.** Invariant $p_T$ spectrum of the single electron (left) and single muon (right) from heavy flavor semi-lepton decay in p+p collisions.

Experimentally, we use pp as a reference to study nuclear medium effects. In d+Au collisions, parton shadowing, color glass condensate, initial state energy loss, and coherent multiple scattering in final state interactions can be studied. The non-photonic single electron production at middle rapidity follows binary scaling within experimental uncertainty (See Figure 2), however the prompt single muon production at forward shows suppression. It is consistent with CGC and power correction pictures. The enhancement at backward direction needs more theoretical investigation. Anti-shadowing and recombination could lead to such enhancement (See Figure 3).

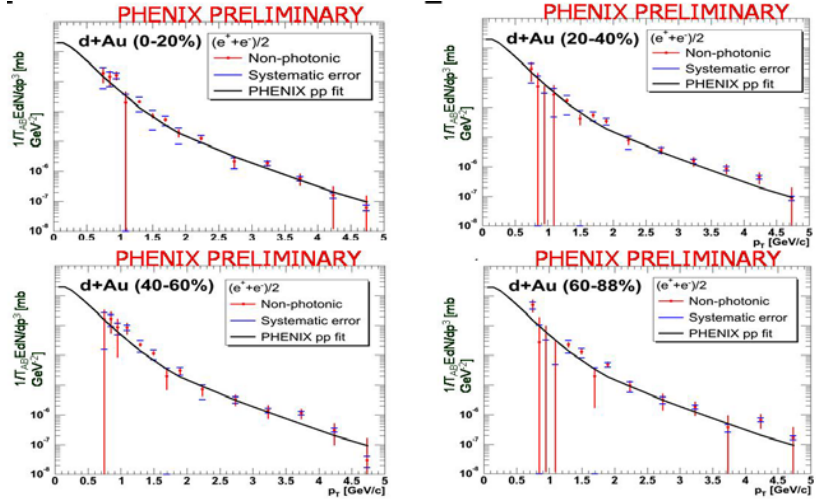

**Figure 2.** Invariant $p_T$ spectrum of the non-photonic electron in different centrality dAu collisions.

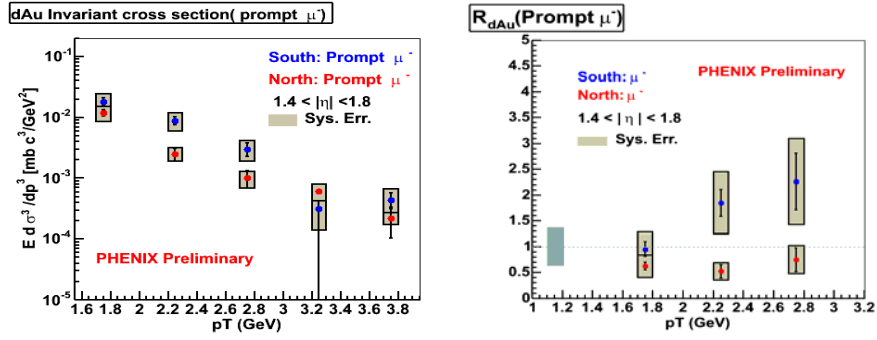

**Figure 3.** Invariant $p_T$ spectra of the prompt muon (left) and $R_{dAu}$ (right) dAu collisions. South Arm is Au-going direction and North Arm is deuteron-going direction.

PHENIX also measured the nuclear modification factor $R_{AA}$ in Au+Au Collisions for the non-photonic electrons mainly come from open heavy flavor decays [6]. Heavy quarks were expected to experience smaller energy loss in the hot dense matter due to a suppression of the phase space for gluon radiation for large masses of the quarks (the so called "dead cone" effect [7]).

The subtracted ``non-photonic'' electron yields for different centrality bins are shown in Figure 4 (left). A fit to the pp data, scaled by corresponding number of binary collisions, is also shown on the plot. One can clearly see that for the most peripheral bin the AuAu data agrees with the pp reference and that a high-$p_T$ suppression of the electron yield starts developing towards more central collisions. The nuclear modification factor $R_{AA}$ for the most central bin is shown on (Figure 4 right). The comparison with the published theoretical predictions [8,9] indicates that a large transport coefficient or a very high initial gluon density in Au+Au collisions.

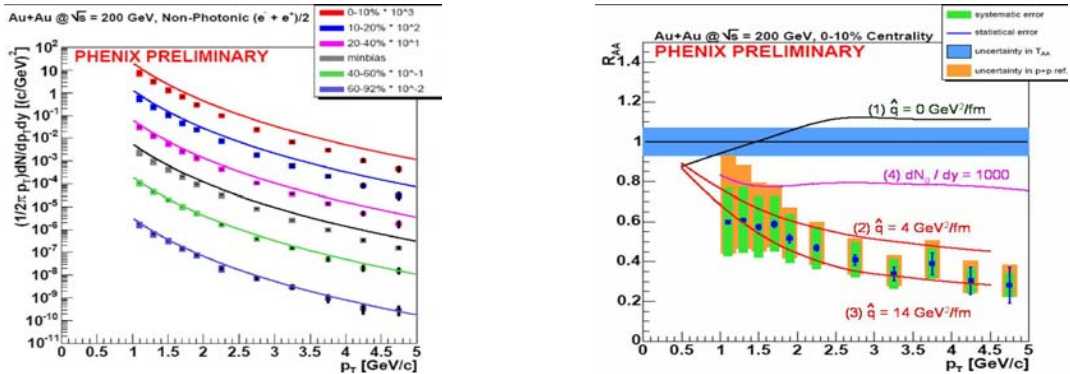

**Figure 4.** Invariant $p_T$ spectra of non-photonic electron (left) and $R_{AA}$ (right) Au+Au collisions.

# REFERENCE

[1] K. Adcox et al., Nucl. Instrum. Methods A499, 469 (2003).
[2] S.S. Adler et al, PRL88, 192303 (2002).
[3] S.S. Adler et al, PRL94, 082301 (2005).
[4] S.S. Adler et al, hep-ex/0508034.
[5] Y. Kwon for PHENIX collaborations, nucl-ex/0510011.
[6] S. Butsyk for PHENIX collaborations, nucl-ex/0510010
[7] Y.L. Dokshitzer and D.E. Kharzeev, PLB519, 199 (2001).
[8] N. Armesto, S. Dainese, C. Salgado, and U. Wiedemann, PRD 71, 054027 (2005).
[9] M. Djordjevic, M. Gyulassy, S. Wicks, PRL 94, 112301 (2005).